\documentclass[11pt,fleqn]{article}
\usepackage{amsfonts,amssymb,latexsym,cite}
\usepackage{epsf,graphicx}

\newcommand{\bls}[1]{\renewcommand{\baselinestretch}{#1}}

\def\noi{\noindent}

\newcommand{\Title}[1]{\noi {\uppercase{\Large #1}}\\[1ex]}

\def\Aunames#1{\noi{\large\bf #1}}
\def\auth#1{${}^{#1}$}
\def\Addresses#1{\medskip\noi \protect
    \begin{description}\itemsep -3pt {\it #1} \end{description}}
\def\addr#1#2{\item[${}^{#1}$]{\it #2}}

\newcommand{\Abstract}[1]{\vskip 2mm \begin{center}
        \parbox{16.4cm}{\small\noi #1} \end{center}\medskip}
\newcommand{\PACS}[1]{\begin{center}{\small PACS: #1}\end{center}}

\def\email#1#2{\footnotetext[#1]{e-mail: #2}\addtocounter{footnote}{1}}

\def\nq{\hspace*{-1em}}
\def\nqq{\hspace*{-2em}}
\def\nhq{\hspace*{-0.5em}}

\def\cm{\hspace*{1cm}}
\def\inch{\hspace*{1in}}

\def\ten#1{\mbox{$\times 10^{#1}$}}

\def\Jl#1#2{#1 {\bf #2},\ }

\def\ApJ#1 {\Jl{Astroph. J.}{#1}}
\def\CQG#1 {\Jl{Class. Quantum Grav.}{#1}}
\def\DAN#1 {\Jl{Dokl. AN SSSR}{#1}}
\def\GC#1 {\Jl{Grav. \& Cosmol.}{#1}}
\def\GRG#1 {\Jl{Gen. Rel. Grav.}{#1}}
\def\JETF#1 {\Jl{Zh. Eksp. Teor. Fiz.}{#1}}
\def\JETP#1 {\Jl{Sov. Phys. JETP}{#1}}
\def\JHEP#1 {\Jl{JHEP}{#1}}
\def\JMP#1 {\Jl{J. Math. Phys.}{#1}}
\def\NPB#1 {\Jl{Nucl. Phys.}{B\ #1}}
\def\NP#1 {\Jl{Nucl. Phys.}{#1}}
\def\PLA#1 {\Jl{Phys. Lett.}{#1A}}
\def\PLB#1 {\Jl{Phys. Lett.}{#1B}}
\def\PRD#1 {\Jl{Phys. Rev.}{D\ #1}}
\def\PRL#1 {\Jl{Phys. Rev. Lett.}{#1}}

\def\al{&\nhq}
\def\lal{&&\nqq {}}
\def\eq{Eq.\,}
\def\eqs{Eqs.\,}
\def\beq{\begin{equation}}
\def\eeq{\end{equation}}
\def\bear{\begin{eqnarray}}
\def\bearr{\begin{eqnarray} \lal}
\def\ear{\end{eqnarray}}
\def\earn{\nonumber \end{eqnarray}}
\def\nn{\nonumber\\ {}}
\def\nnv{\nonumber\\[5pt] {}}
\def\nnn{\nonumber\\ \lal }
\def\nnnv{\nonumber\\[5pt] \lal }
\def\yy{\\[5pt] {}}
\def\yyy{\\[5pt] \lal }
\def\eql{\al =\al}

\def\dst{\displaystyle}

\def\fracd#1#2{{\dst\frac{#1}{#2}}}

\def\Half{{\fracd{1}{2}}}

\def\e{{\,\rm e}}
\def\d{\partial}

\def\sign{\mathop{\rm sign}\nolimits}

\def\eps{\varepsilon}

\def\then{\ \Rightarrow\ }

\def\mn{_{\mu\nu}}
\def\MN{^{\mu\nu}}
\def\mN{_\mu^\nu}
\def\nM{_\nu^\mu}
\def\cK{{\cal K}}
\def\cV{{\cal V}}

\def\kappa{\varkappa}

\def\wt{\widetilde}
\def\tg{{\wt g}}
\def\tR{{\wt R}}

\def\H{{\mathbb H}}

\def\oR{{\overline R}}

\def\sss{\scriptscriptstyle}
\def\aE{a_{\sss\rm E}{}}
\def\dotaE{{\dot a}_{\sss\rm E}{}}
\def\mD{m_{\sss D}}

\bls{0.98}
\begin{document}
\thispagestyle{empty}
\twocolumn[

\Title {Cosmologies from nonlinear multidimensional gravity\yy
    	with acceleration and slowly varying $G$}

\Aunames
{K. A. Bronnikov\auth{a,b,1}, S. A. Kononogov\auth{a,2},
    V. N. Melnikov\auth{a,b,3}, and S. G. Rubin\auth{c,4}}

\Addresses{
    \addr a {Center for Gravitation and Fundamental Metrology,
        VNIIMS, 46 Ozyornaya St., Moscow 119361, Russia }
    \addr b {Institute of Gravitation and Cosmology,
        Peoples' Friendship University of Russia,\\
        6 Miklukho-Maklaya St., Moscow 117198, Russia}
    \addr c {Moscow State Engineering Physics Institute,
        31 Kashirskoe Sh., Moscow 115409, Russia}
   }

\Abstract{We consider multidimensional gravity with a Lagrangian containing
  the Ricci tensor squared and the Kretschmann invariant. In a Kaluza-Klein
  approach with a single compact extra space of arbitrary dimension,
  with the aid of a slow-change approximation (as compared with the Planck
  scale), we build a class of spatially flat cosmological models in which
  both the observed scale factor $a(\tau)$ and the extra-dimensional one,
  $b(\tau)$, grow exponentially at large times, but $b(\tau)$ grows slowly
  enough to admit variations of the effective gravitational constant $G$
  within observational limits. Such models predict a drastic change in the
  physical laws of our Universe in the remote future due to further growth
  of the extra dimensions.

  }
\vspace{-4mm}

] 
\email 1 {kb20@yandex.ru}
\email 2 {kononogov@vniims.ru}
\email 3 {melnikov@phys.msu.ru}
\email 4 {sergeirubin@list.ru}

\section{Introduction}

  Multidimensional models of gravity and cosmology as low-energy limits
  of unified theories of physical interactions \cite{Mel1} are a powerful
  tool for studies of the present challenges to modern physics
  \cite{Solv,Paris,Mel-07}, such as the dark energy and dark matter
  problems, possible variations of fundamental constants \cite{KM,BK,BKM},
  the role of strong field objects (black holes, wormholes) etc.

  It has been recently argued \cite{VfromD,brost,asym_bw} that
  multidimensional gravity with curvature-nonlinear terms in the action can
  be a source of a great diversity of effective theories able to address a
  number of important problems of modern astrophysics and cosmology using a
  minimal set of postulates. Among such problems one can mention the essence
  of dark energy, early formation of supermassive black holes (which is a
  necessary stage in some scenarios of cosmic structure formation), and
  sufficient particle production at the end of inflation. In this approach,
  the particular value of the total space-time dimension $D > 4$ and the
  topological properties of space-time are supposed to be determined by
  quantum fluctuations and may vary from one space-time region to another,
  leading to drastically different universes. Different effective theories
  can take place even with fixed parameters of the original Lagrangian.
  It can be shown that this approach, without need for fields other than
  gravity, is able to produce such different structures as inflationary
  (or simply accelerating) universes, brane worlds \cite{asym_bw}, black
  holes etc. The role of scalar fields is played by the metric components
  of extra dimensions.

  One of the challenging problems of modern physics and cosmology is that of
  possible time-, location-, and scale-dependent variations of the
  fundamental physical constants, in particular, of Newton's gravitational
  constant $G$.  Variable effective constants are a common feature of
  multidimensional cosmologies, where these constants depend on the
  properties of extra dimensions which can vary from one space-time point to
  another. This problem, among others, was addressed in \cite{VfromD}. A
  number of examples of cosmological models were built, where the size of
  the extra dimensions was stabilized at a minimum of the effective
  potential. In such models, the constants could vary at earlier stages
  while the effective scalar field only approached this minimum, but take on
  stable values together with the scalar when this minimum is achieved.

  In the present paper, we discuss another type of multidimensional
  cosmologies (their possibility was also mentioned in \cite{VfromD}) in
  which the extra-dimensional scale factor can grow indefinitely at large
  times but is yet small enough at present. Such models are of interest
  since they predict rather an exotic, though sad, fate of the Universe we
  live in: sooner or later, its physical laws must drastically change due to
  the growth of the internal space. We will show that such models can in
  principle be viable since they simultaneously predict an accelerated (de
  Sitter-like) expansion and, under certain conditions on the input
  parameters of the theory, a small enough variation of the gravitational
  constant in agreement with observations.

\section {The multidimensional theory and its reduction}

  We consider a $(D = 4 + d_1)$-dimensional manifold with the metric
\beq                                                      \label{ds}
    ds^2 = g\mn dx^\mu dx^\nu + \e^{2\beta(x)} b_{ab} dx^a dx^b
\eeq
  where the extra-dimensional metric components $b_{ab}$ are independent of
  $x^{\mu}$, the observable four space-time coordinates.

  The $D$-dimensional Riemann tensor has the nonzero components
\bear
    R\MN{}_{\rho\sigma} \eql \oR\MN{}_{\rho\sigma},       \label{Riem}
\nn
    R^{\mu a}{}_{\nu a} \eql \delta^a_b\, B\nM, \cm
    B\nM := \e^{-\beta} \nabla_\nu (\e^\beta \beta^{\mu}),        
\nn
    R^{ab}{}_{cd} \eql
      \e^{-2\beta} \oR^{ab}{}_{cd} + \delta^{ab}{}_{cd} \beta_\mu\beta^\mu,
\ear
  where capital Latin indices cover all $D$ coordinates, the bar marks
  quantities obtained from $g\mn$ and $b_{ab}$ taken separately,
  $\beta_{\mu} \equiv \d_{\mu}\beta$ and $\delta^{ab}{}_{cd}\equiv
  \delta_{c}^{a}\delta_{d}^{b}-\delta_{d}^{a}\delta_{c}^{b}$. The
  nonzero components of the Ricci tensor and the scalar curvature are
\bear
    R\mN \eql \oR\mN + d_1\, B\mN,          \label{Ric}           
\nnv
    R_a^b \eql \e^{-2\beta} \oR_a^b
                    + \delta_a^b [ \Box \beta + d_1 (\d{\beta})^2 ],
\nnv
    R \eql \oR [g] + \e^{-2\beta }\oR [b] + 2d_1 \Box \beta
\nnn \cm \cm
           + d_1(d_1+1) (\d{\beta})^2,
\ear
  where $(\d{\beta})^2 \equiv \beta_{\mu}\beta^{\mu}$,
  $\Box = \nabla^\mu \nabla_\mu$ is the d'Alembert operator while $\oR[g]$
  and $\oR [b]$ are the Ricci scalars corresponding to $g\mn$ and $b_{ab}$,
  respectively. Let us also present, using similar notations, the
  expressions for two more curvature invariants, the Ricci tensor squared
  and the Kretschmann scalar $\cK = R^{ABCD}R_{ABCD}$:
\bearr
    R_{AB}R^{AB} = \oR\mn\oR\MN + 2d_1 \oR \mn B\MN
        + \e^{-4\beta}\oR_{ab}\oR^{ab}
\nnn\cm
    + 2\e^{-2\beta} \oR[h] [\Box\beta + d_1 (\d{\beta})^2 ]
\nnn\cm
      + d_1 [\Box\beta + d_1 (\d{\beta})^2 ]^2               \label{Ric2}
\yyy
    \cK = \overline{\cK}[g] + 4 d_1 B\mn B\MN + \e^{-4\beta}\overline{\cK}[h]
\nnn \quad\
      + 4 \e^{-2\beta} \oR [h] (\d{\beta})^2
         +2 d_1 (d_1-1) [(\d\beta)^2 ]^2 .                     \label{Kre}
\ear

  Suppose now that $b_{ab}$ describes a compact $d_1$-dimensional space of
  nonzero constant curvature, i.e., a sphere ($k=1$) or a compact
  $d_1$-dimensional hyperbolic space \cite{Lob} ($k = -1$) with a
  fixed curvature radius $r_0$ normalized to the $D$-dimensional
  analogue $\mD$ of the Planck mass, i.e., $r_0 = 1/\mD$ (we use the
  natural units, with the speed of light $c$ and Planck's constant $\hbar$
  equal to unity). We have
\bear
    \oR^{ab}{}_{cd} \eql k\, \mD^2\,\delta^{ab}{}_{cd},         \label{r0}
\nn
    \oR_a^b \eql k\, \mD^2\, (d_1-1) \delta_a^b,
\nn
    \oR [b] \eql k\, \mD^2\, d_1 (d_1-1) = R_b.
\ear
  The scale factor $b(x) \equiv \e^{\beta}$ in (\ref{ds}) is thus kept
  dimensionless; $R_b$ has the meaning of a characteristic curvature scale
  of the extra dimensions.

  Consider, in the above geometry, a sufficiently general
  curvature-nonlinear theory of gravity with the action
\bear                                                         \label{act1}
     S \eql \Half \mD^{D-2} \int\sqrt{^{D}g}\,d^{D}x\,(L_g + L_m),
\nn
     L_g \eql F(R) + c_1 R^{AB}R_{AB} + c_2 \cK,
\ear
  where $F(R)$ is an arbitrary smooth function, $c_1$ and $c_2$ are
  constants, $L_m$ is a matter Lagrangian and ${^D}g = |\det(g_{MN})|$.

  The extra coordinates are easily integrated out, reducing the action
  to four dimensions:
\beq
     S = \Half \cV [d_1]\,\mD^2 \int\sqrt{^4g}\,d^{4}x\,
            \e^{d_1\beta}\,[L_g + L_m],                      \label{act2}
\eeq
  where $^{4}g = |\det(g\mn)|$ and $\cV[d_1 ]$ is the volume of a
  compact $d_1$-dimensional space of unit curvature.

  \eq (\ref{act2}) describes a curvature-nonlinear theory with non-minimal
  coupling between the effective scalar field $\beta$ and the curvature.
  Let us simplify it in the following way (putting, for convenience, $\mD=1$):

\medskip\noi
{\bf (a)} Express everything in terms of 4D
    variables and $\beta(x)$; we have, in particular,
\bearr                                                      \label{R4}
        R = R_4 + \phi + f_1,
\nnnv
        R_4 = \oR [g],      \quad\
            f_1 = 2d_1 \Box \beta + d_1(d_1+1)(\d{\beta})^2,
\ear
    where we have introduced the effective scalar field
\beq                                                    \label{phi}
        \phi (x) = R_b \e^{-2\beta (x)}
              = k d_1(d_1-1)\, \e^{-2\beta (x)}.
\eeq
    The sign of $\phi$ coincides with $k = \pm 1$, the sign of curvature in
    the $d_1$ extra dimensions.

\medskip\noi
{\bf (b)} Suppose that all quantities are slowly varying, i.e., consider
    each derivative $\d_{\mu}$ (including those in the definition of $\oR$)
    as an expression containing a small parameter $\eps$; neglect all
    quantities of orders higher than $O(\eps^2 )$ (see \cite{VfromD,Don}).

\medskip\noi
{\bf (c)} Perform a conformal mapping leading to the Einstein conformal
    frame, where the 4-curvature appears to be minimally coupled to the
    scalar $\phi$.

\medskip
    In the decomposition (\ref{R4}), both terms $f_1$ and $R_4$ are regarded
    small in our approach, which actually means that all quantities,
    including the 4D curvature, are small as compared with the
    $D$-dimensional Planck scale. The only term which is not small is
    $\phi$, and we can use a Taylor decomposition of the function $F(R) =
    F(\phi + R_4 + f_1)$:
\bearr                                                       \label{Fapprox}
    F(R) = F(\phi + R_4 + f_1 )
\nnn \cm
    \simeq F(\phi) + F'(\phi)\cdot(R_4 +f_1 )+...,
\ear
    with $F'(\phi)\equiv dF/d\phi$. Substituting this, and the corresponding
    decompositions of the expressions (\ref{Ric2}) and (\ref{Kre}), into \eq
    (\ref{act2}), we obtain, up to $O(\eps^2)$, the following effective
    gravitational Lagrangian $L_g$ in \eq (\ref{act2}):
\bearr
      L_g = F'(\phi) R_4  + F(\phi) + F'(\phi) f_1 + c_*\phi^2
\nnn \cm
      + 2 c_1\phi \Box\beta
            + 2(c_1 d_1 + 2c_2) (\d\beta)^2                \label{Lg_4}
\ear
    with  $c_* = c_1/d_1 + 2c_2/[d_1(d_1-1)]$.

    The action (\ref{act2}) with (\ref{Lg_4}) is typical of a scalar-tensor
    theory (STT) of gravity in a Jordan frame. To study the dynamics of the
    system, it is helpful to pass on to the Einstein frame. Applying the
    conformal mapping
\bearr  \nhq                                                \label{trans-g}
    g\mn \ \mapsto \tg\mn = |f(\phi)| g\mn,
\quad
            f(\phi) =  \e^{d_1\beta}F'(\phi),
\ear
    after a lengthy calculation, we obtain the action in the Einstein frame
    as
\bear
     S \eql \Half \cV[d_1] \int \sqrt{\tg}\, (\sign F') L,
\nn
     L \eql \tR_4 + \Half K_{\rm E}(\phi) (\d\phi)^2
                        - V_{\rm E}(\phi) + {\wt L}_m,      \label{Lgen}
\\
     {\wt L}_m \eql (\sign F')\frac{\e^{-d_1\beta}}{F'(\phi)^2} L_m;
                                 \label{Lm}
\\ \nq
     K_{\rm E}(\phi) \eql                                   \label{KE}
        \frac{1}{2\phi^2} \biggl[
            6\phi^2 \biggl(\frac{F''}{F'}\biggr)^2\!
            -2 d_1 \phi \frac{F''}{F'}
\nnn \cm
        + \Half d_1 (d_1{+}2) + \frac{4(c_1 + c_2)}{F'}\biggr],
\\ \nq
     V_{\rm E}(\phi) \eql - (\sign F') \frac{\e^{-d_1\beta}}{F'(\phi)^2}
                [F(\phi) + c_* \phi^2],
                                  \label{VE}
\ear
    where the tilde marks quantities obtained from or with $\tg\mn$;
    the indices are raised and lowered with $\tg\mn$; everywhere $F =
    F(\phi)$ and $F' = dF/d\phi$; $\e^{\beta}$ is expressed in terms of
    $\phi$ using (\ref{phi}).

\section {The cosmological model}

    Depending on the choice of $F(R)$, the parameter $c_1$ and $c_2$ and
    the matter Lagrangian in the action (\ref{act1}), the theory under
    consideration can lead to a great variety of cosmological models. Many
    of them were discussed in \cite{VfromD}, mostly those related to minima
    of the effective potential (\ref{VE}) at nonzero values of $\phi$. Such
    minima correspond to stationary states of the scalar $\phi$ and
    consequently of the scale factor $b = \e^{\beta}$ of the extra
    dimensions (see also \cite{Zhuk}).  If the minimum value of the
    potential is positive, it can play the role of a cosmological constant
    that launches an accelerated expansion of the Universe.

    Here, we would like to pay attention to one more minimum of the
    potential $V_{\rm Ein}$, existing for generic choices of the function
    $F(R)$ with $F' >0$ and located at the point $\phi =0$. The asymptotic
    $\phi \to 0$ corresponds to growing rather than stabilized extra
    dimensions: $b = \e^{\beta}\sim 1/\sqrt{|\phi|} \to \infty$. A model
    with such an asymptotic growth at late times may still be of interest if
    the growth is sufficiently slow and the size $b$ does not reach
    detectable values by now. Let us recall that the admissible range of
    such growth comprises as many as 16 orders of magnitudes if the
    $D$-dimensional Planck length $1/\mD$ coincides with the 4D one, i.e.
    about $10^{-33}$ cm. This estimate certainly changes if there is no such
    coincidence.

    Let us check whether it is possible to describe the modern state of
    the Universe by an asymptotic form of the solution for $\phi\to 0$
    as a spatially flat cosmology with the 4D Einstein-frame metric
\beq
        d{\wt s}{}^2_4 = dt^2 - \aE^2 (t) d\vec x{}^2,         \label{dsE}
\eeq
    where $\aE$ is the Einstein-frame scale factor. At small $\phi$,
    assuming a smooth function $F(\phi)$, we can use its Taylor
    decomposition
\beq                                                         \label{F_0}
    F(\phi) = -2\Lambda + \phi + c\phi^2 +\ldots,
\eeq
    where the form of the first two terms is chosen to obtain in
    (\ref{act1}) multidimensional Einstein gravity in the first
    approximation in $R$; $\Lambda$ is the initial cosmological constant.
    For the kinetic and potential terms in the Lagrangian (\ref{Lgen}) we
    then have in the main approximation with respect to $\phi$:
\bearr
        K_{\rm E} \approx K_0/\phi^2,
\nnn    \cm                                \label{KV_0}
        K_0 = \frac{1}{4}[d_1 (d_1 +2) + 8(c_1+c_2)];
\nnn
    V_{\rm E} \approx \Lambda \e^{-d_1\beta}
            = \Lambda [d_1(d_1-1)]^{d_1/2} |\phi|^{d_1/2}.
\ear

    Assuming $L_m = 0$ (i.e., restricting ourselves to vacuum cosmologies),
    we can write two independent components of the Einstein-scalar equations
    for $\beta(t)$ and $\aE(t)$ in the form
\bear                                                        \label{eq-b4}
       \ddot \beta + 3\frac{\dotaE}{\aE} \dot{\beta}
                \eql \frac{\Lambda d_1} {4K_0} \e^{-d_1\beta},
\\
       3\frac{\dotaE^2}{\aE^2}                                \label{eq-a4}
            \eql 2K_0 \dot\beta{}^2
                    + 6\Lambda \e^{-d_1\beta}.
\ear

    It is hard to solve this set of equations exactly.
    Let us, however, notice that our equations are of the same type as those
    appearing in studies of inflationary cosmologies, and the field $\beta$
    plays the role of an inflaton. Hence it seems possible to apply the
    slow-rolling approximation frequently used there: we suppose
\beq                                                       \label{rolling}
       |\ddot \beta| \ll 3\frac{\dotaE}{\aE} \dot{\beta},
   \qquad
        K_0 \dot\beta{}^2 \ll 3\Lambda \e^{-d_1\beta}.
\eeq
    and drop the corresponding terms in \eqs (\ref{eq-a4}) and
    (\ref{eq-b4}). Then we can express $\dotaE/\aE$ from (\ref{eq-a4}) and
    insert it to (\ref{eq-b4}), getting
\beq                                                           \label{eq-b}
    \dot\beta = \frac{d_1^2 \sqrt{2\Lambda}}{24 K_0}\e^{-d_1\beta/2},
\eeq
    which is easily integrated to give
\beq                                                           \label{sol-b}
    \e^{d_1\beta/2} = \frac{d_1^2\sqrt{2\Lambda}}{48 K_0}(t-t_0),
\eeq
    where $t_0$ is an integration constant which can be eliminated without
    loss of generality by changing the zero point of the coordinate $t$.
    For the scale factor $\aE$ we obtain
\beq                                                           \label{sol-aE}
     \dotaE/\aE = p/t \quad\then\quad \aE \propto t^p
\eeq
    with
\beq
    p = 48 K_0/d_1^2.                                  \label{p}
\eeq

    Substituting the solution to the slow-rolling conditions
    (\ref{rolling}), we see that both of them hold if $3p \gg 1$, or, in
    terms of the input parameters of the theory,
\beq                                                         \label{roll}
    3p = 36 \biggl[\frac{d_1+2}{d_1} + \frac{8(c_1+c_2)}{d_1^2}\biggr]\gg 1.
\eeq
    For $c_1 + c_2 \geq 0$, we have $3p > 36$. This verifies a sufficiently
    good precision of our solution.

    Passing over to the Jordan frame with
\beq                                                        \label{ds_j}
      ds^2_4 = d\tau^2 - a^2(\tau) d\vec x{}^2
        = \frac{1}{f}[ dt^2 - \aE^2(t) d\vec x{}^2]
\eeq
    (so that $\tau$ is the cosmic time in the Jordan frame), due to $F'(0)
    \approx 1$, we can put simply $f = \e^{d_1\beta} \approx t^2$, to obtain
\bearr                                                        \label{jord}
            t = \e^{\sqrt{2\Lambda}\tau/p},\cm
     a(\tau) \propto \e^{(p-1)\sqrt{2\Lambda}\tau/p},
\nnn
     b(\tau) = \e^{\beta(\tau)} = \left(\frac{\sqrt{2\Lambda}}{p}
                  \e^{\sqrt{2\Lambda}\tau/p}\right)^{2/d_1},
\ear
    where we have fixed an integration constant by choosing the zero point
    of the time variable $\tau$.

    A further interpretation of the results depends on which conformal
    frame is regarded physical (observational) \cite{bm-predict,erice},
    and this in turn depends on the manner in which fermions appear in the
    (so far unknown) underlying unification theory involving all
    interactions. We here restrict ourselves to the simplest and maybe the
    most natural assumption, that the observational frame coincides with the
    fundamental (Jordan) one, in which the initial theory (\ref{act1}) has
    been formulated.

    Then an immediate observation is that the external scale factor
    $a(\tau)$ grows exponentially, in a de Sitter manner, which conforms to
    modern observations if one properly chooses the constants, namely,
\bearr
       (p-1)\sqrt{2\Lambda}/p = H_0 \approx 2.3\ten{-18}\ {\rm s}^{-1}
\nnn \inch
                \approx 7.25\ten{-11}\ {\rm yr}^{-1},  \label{H0}
\ear
    where $H_0$ is the modern value of the Hubble parameter.

    The internal scale factor $b(\tau)$ grows much slower for sufficiently
    large $d_1$, while the volume factor $b^{d_1}$ behaves like
    $\e^{2\sqrt{2\Lambda}\tau/p}$. The effective gravitational constant is
    known to change inversely proportionally to the volume factor, so that
\beq
    {\dot G}/G = -2\sqrt{2\Lambda}/p.
\eeq
    (here and henceforth the dot means $d/d\tau$).
    The dimensionless parameter of $G$ variation is
\beq                                                       \label{delta}
    \delta = {\dot G}/(GH_0) = -2/(p-1).
\eeq

   The tightest experimental constraint on $G$ variation has been obtained,
   to our knowledge, from lunar laser ranging (LLR) data \cite{muller07},
   namely,
\beq
    {\dot G}/G = (2 \pm 7)\ten{-13}\ {\rm yr}^{-1}.
\eeq
   Thus a viable cosmology should predict $\delta \lesssim 10^{-2}$,
   which according to (\ref{delta}) constrains our model parameters by
\beq
     p = 12 \biggl[ \frac{d_1+2}{d_1} + \frac{8(c_1+c_2)}{d_1^2}\biggr]
    \gtrsim 100,                                           \label{p-1}
\eeq

   It follows that our initial field model (\ref{act1}) with only $F(R)$ is
   unable to satisfy the constraint (\ref{p-1}): to do so, it is necessary
   to invoke the Ricci tensor squared or/and the Kretschmann scalar, with
   the input constants $c_1$ and $c_2$ such that
\beq
    c_1 +c_2 \gtrsim d_1^2.                               \label{c1,2}
\eeq

   We conclude that, under the condition (\ref{c1,2}), our
   model with the asymptotic behavior (\ref{jord}) is potentially viable
   since it combines a de Sitter-like expansion of the observable Universe
   with a slow enough variation of the effective gravitational constant.

   Three more observations can be added. First, comparing the constraints
   (\ref{p-1}) and (\ref{roll}), we see that our approximation works so much
   the better, the smaller is variation of $G$.

   Second, according to (\ref{H0}), the constant $\Lambda$ is determined by
   the modern value of the Hubble parameter and is thus approximately the
   same as the observed cosmological constant. So this model suffers the
   same fine-tuning problem for the cosmological constant value as the
   Standard model.

   Third, the size $b(\tau)$ of the extra dimensions, given by (\ref{jord}),
   remains uncertain since it depends on the arbitrarily chosen values of
   the time variable $\tau$ (note that observable quantities depend on $H_0$
   rather than $\tau$). Thus one can easily satisfy the conditions that
   $b(\tau)$ must be much greater than the Planck length but be still below
   the observational threshold (about $10^{-17}$ cm).
   We also see that even an exponential growth of the extra dimensions can
   be slow enough to conform to the observational bounds on the stability
   of fundamental constants.

\medskip\noi
{\bf Acknowledgment.} This work was supported in part by the Russian Basic
Research Foundation grant 07-02-13624-ofi-ts.


\small
\begin{thebibliography}{99}

\bibitem{Mel1}
	V.N. Melnikov, {\it Multidimensional Classical and Quantum
	Cosmology and Gravitation. Exact Solutions and Variations of
	Constants\/}, CBPF-NF-051/93, Rio de Janeiro, 1993;
	in: {\it Cosmology and Gravitation\/}, ed. M. Novello
		(Editions Frontieres, Singapore, 1994), p. 147;\\
	{\it Multidimensional Cosmology and  Gravitation\/},
	CBPF-MO-002/95, Rio de Janeiro, 1995, 210 p.;
	in: {\it Cosmology and Gravitation. II}, ed. M. Novello
		(Editions Frontieres, Singapore, 1996), p. 465;\\
	{\it Exact Solutions in Multidimensional Gravity and Cosmology
	III\/}, CBPF-MO-03/02,  Rio de Janeiro, 2002, 297 pp.

\bibitem {Solv}
	V.N. Melnikov, {\it Gravity as a key problem of the millennium\/}.
	Proc. 2000 NASA/JPL Conference on Fundamental Physics in
	Microgravity, NASA Document D-21522, 2001, p. 4.1--4.17, Solvang,
	CA, USA.

\bibitem{Paris}
	V.N. Melnikov, {\it Gravity and cosmology as key problems of the
	millennium\/}. In: {\it Proc. Albert Einstein Century Int. Conf.\/},
	eds. J.-M. Alimi and A. Fuzfa (AIP Conf. Proc., Melville--NY,
	2006), v. 861, p. 109--126.

\bibitem{Mel-07}
 	V.N. Melnikov,  Grav. Cosmol. {\bf 13}, 81 (2007).

\bibitem{KM}
	S.A. Kononogov and V.N. Melnikov,
		Izmeritel'naya Tekhnika {\bf 6}, 1 (2005).

\bibitem{BK}
	K.A. Bronnikov and S.A. Kononogov, Metrologia {\bf 43}, R1 (2006).

\bibitem{BKM}
	K.A. Bronnikov,	S.A. Kononogov and V.N. Melnikov,
		\GRG {38} 1215 (2006).

\bibitem{VfromD}
    	K.A. Bronnikov and S.G. Rubin, \PRD {73} 124019 (2006).

\bibitem{brost}
    	K.A. Bronnikov, R.V. Konoplich and S.G. Rubin,
        \CQG {24} 1261 (2007).

\bibitem{asym_bw}
    	K.A. Bronnikov and S.G. Rubin, \GC {13} 191 (2007).

\bibitem{Lob}
       	Compact hyperbolic spaces of constant curvature on the basis of a
        usual open Lobachevsky space $\H^d$ are isometric to quotient spaces
        $\H^d/\Gamma$ where $\Gamma$ is a nontrivial discrete group of
        isometries of $\H^d$, see, e.g., B.A. Dubrovin, A.T. Fomenko and S.P.
        Novikov, {\it Modern Geometry --- Methods and Applications}
        (Springer-Verlag, New York, 1992). On possible applications of such
        (3D) spaces in cosmology see, e.g., D. M\"uller, H.V. Fagundes and
        R. Opher, \PRD {66} 083507 (2002) and references therein.

\bibitem{Don}
     	J.F. Donoghue, \PRD {50} 3874 (1994).

\bibitem{Zhuk}
     	U. G\"unther, P. Moniz and A. Zhuk, Astrophys. Space Sci.
		{\bf 283}, 679-684 (2003); gr-qc/0209045;\\
     	U. G\"unther and A. Zhuk, {\it Remarks on dimensional reduction in
		multidimensional cosmological models\/}, gr-qc/0401003.

\bibitem{bm-predict}
     	K.A. Bronnikov and V.N. Melnikov, \GRG {33} 1549 (2001).

\bibitem{erice}
     	K.A. Bronnikov and V.N. Melnikov,
        ``Conformal frames and D-dimensional gravity'',
        gr-qc/0310112,
     	in: {\it Proc. 18th Course of the School on Cosmology and
        Gravitation: The Gravitational Constant. Generalized Gravitational
        Theories and Experiments\/} (30 April--10 May 2003, Erice),
        Ed. G.T. Gillies, V.N. Melnikov and V. de Sabbata,
        (Kluwer, Dordrecht/Boston/London, 2004) pp. 39--64.

\bibitem{muller07}
    	J. M\"uller and L. Biskupek, \CQG {24} 4533 (2007).

\end{thebibliography}
\end{document}